# Degradation of methylparaben by anodic oxidation, electro-Fenton, and photoelectro-Fenton using carbon felt-BDD cell

Aline B. Trench[1,2,*], Nihal Oturan[1], Aydeniz Demir[3], João P. C. Moura[2], Clément Trellu[1], Mauro C. Santos[2,*], Mehmet A. Oturan[1,*]

[1]*Laboratoire Géomatériaux et Environnement EA 4508, Université Gustave Eiffel, 5 Bld Descartes, 77454 Marne-la-Vallée, Cedex 2, France.*

[2]*Laboratório de Eletroquímica e Materiais Nanoestruturados (LEMN), Centro de Ciências Naturais e Humanas (CCNH), Universidade Federal do ABC (UFABC). Rua Santa Adélia 166, Bairro Bangu, 09210-170, Santo André - SP, Brasil.*

[3]*Department of Environmental Engineering, Faculty of Engineering, Mersin University, Çiftlikköy Campus, 33343, Mersin, Turkey.*

*Corresponding author,* E-mail address:

aline_trench@hotmail.com (A. B. Trench)

mauro.santos@ufabc.edu.br (M. C. Santos)

mehmet.oturan@univ-eiffel.fr (M. A. Oturan)

**Abstract**

In this study, the comparative efficiency of different electrochemical advanced oxidation processes, such as anodic oxidation with electrogenerated $H_2O_2$ (AO-$H_2O_2$), electro-Fenton (EF), and its combination with UV irradiation (photoelectron-Fenton (PEF)), was investigated for the removal of methylparaben (MP) using a carbon felt cathode and a boron-doped diamond anode. The EF process achieved a higher MP removal efficiency than the AO-$H_2O_2$ process for all applied current densities. The total organic carbon (TOC) removal after 6 h of treatment at a current density of 10 mA cm$^{-2}$ reached 75.0% and 91.9% for the AO-$H_2O_2$ and EF processes, respectively. The combination of EF and UV light improved the efficiency of the EF process. The PEF process achieved a TOC removal of 84.6% in only 2 h at 5 mA cm$^{-2}$ and 96.8% after 6 h of treatment. Furthermore, based on identifying oxidation reaction intermediates and short-chain carboxylic acids generated during the treatment, a reaction pathway for methylparaben mineralization by hydroxyl radicals was proposed.

## 1. Introduction

Methylparaben (MP), a chemical compound in the class of compounds called parabens, has a broad spectrum of antimicrobial agents and antibacterial activity [1]. Furthermore, MP exhibits high solubility, stability across different pH values, and a low cost; it is widely used in the cosmetic, food, and pharmaceutical industries [2, 3]. Due to the excessive use of industrialized products and the low efficiency of conventional water treatment methods, MP has been detected in different water sources [4-7]. For example, it was detected in surface water at a concentration of 0.53 mg $L^{−1}$ in Nigeria [8], up to 2.875 ng $L^{−1}$ in Brazil [9], and 3.62 μg $L^{−1}$ in urban rivers in Australia [10]. The situation becomes more aggravating because MP is classified as an endocrine disruptor, posing risks to living beings [11, 12]. Even at low concentrations, parabens have estrogenic activity in living systems and exhibit the potential to lead to the development of different types of cancer [11, 13-15]. Therefore, developing effective treatment methods to eliminate these parabens from water is extremely important.

There are different wastewater treatment methods, such as physical methods (i.e., membrane filtration [16] and adsorption [17]), biological methods (i.e., phytoremediation [18] and bioaugmentation [19]), chemical methods (i.e., chemical precipitation [20] and chemical coagulation [21]), and electrochemical methods [22, 23]. These methods have limited efficiency in the case of persistent organic pollutants. On the other hand, electrochemical advanced oxidative processes (EAOPs), considered green and environmentally friendly methods, have shown excellent efficiency in wastewater treatment [24-29]. They are based on the *in situ* generation of strong oxidants, such as the hydroxyl radical (•OH), which is a non-selective oxidant capable of mineralizing, i.e., converting a wide range of organic pollutants into $CO_2$, $H_2O$, and inorganic ions [22, 23, 30-32]. This radical can be generated in EAOPs either directly or indirectly. In anodic oxidation (AO), it is generated directly by oxidation of water on an appropriate anode (Eq. (1)). The •OH, thus heterogeneously formed, is adsorbed on the anode surface and denoted (M(•OH)). Its reactivity depends on the nature of the anode material (M) [33, 34].

$$M + H_2O \rightarrow M(•OH) + H^+ + e^- \tag{1}$$

In the electro-Fenton (EF) process, •OH radicals are homogeneously generated in the bulk solution from the Fenton reaction (Eq. (2)), in which the hydrogen peroxide ($H_2O_2$) is *in situ* generated on an appropriate cathode through 2-electron reduction of dissolved oxygen (Eq. (3)). The generated $H_2O_2$ reacts with an externally added catalyst ($Fe^{2+}$) to generate homogeneous •OH (Eq. (2)). The catalyst ($Fe^{2+}$) is regenerated during the process by electrochemical reduction of $Fe^{3+}$ formed in the Fenton reaction, according to Eq. (4) [35, 36]. Thus, the continuous electrogeneration of $H_2O_2$ and electrocatalytic regeneration of $Fe^{2+}$ ions allow the constant formation of •OH in the solution. [37].

$$Fe^{2+} + H_2O_2 \rightarrow Fe^{3+} + •OH + OH^- \tag{2}$$

$$O_{2(g)} + 2\,H^+_{(aq)} + 2\,e^- \rightarrow H_2O_2 \tag{3}$$

$$Fe^{3+} + e^- \rightarrow Fe^{2+} \tag{4}$$

The type of anode used in EAOPs can significantly affect pollutant degradation efficiently. For instance, the boron-doped diamond (BDD) anode is considered the best anode material for the electrooxidation of organic pollutants. It has a high overpotential for the $O_2$ evolution reaction and generates a more significant amount of •OH that is poorly adsorbed on its surface, thereby efficiently degrading organic pollutants [33, 34, 38]. Pueyo *et al*. [39] compared the degradation of butylparaben using BDD, stainless steel, and platinum (Pt) anodes via AO process. The results showed that 100% degradation was achieved in 15 min with the BDD anode at a small concentration of 0.5

mgL$^{-1}$ of butylparaben. In contrast, under the same operating conditions, only 40% and 18% degradation were obtained using the stainless steel and Pt anodes. In another study, Mbaye *et al*. [40] compared the use of BDD and Pt anodes in the EF process for electrochemical degradation of the fungicide thiram. A mineralization rate of 92% was achieved for the removal of thiram at the initial concentration of 19.2 mg L$^{-1}$ using the BDD anode. In contrast, the Pt anode degraded only 68.2% under the same conditions. Can *et al*. [41] reported a complete mineralization of bisphenol A through AO using BDD anode. Other anodes tested, i.e., Pt and mixed metal oxide (MMO), showed 60% and 54% degradation rates, respectively, demonstrating the greater efficiency of the BDD anode. Oturan *et al*. [42] compared the efficiency of ten electrodes (including five anodes) for removing the para-aminosalicylic acid. They confirmed the superiority of the BDD anode in the electro-Fenton process. Therefore, EOAPs, such as AO and EF, are considered promising for degrading organic pollutants with BDD anodes.

Using an efficient cathode for the *in-situ* generation of $H_2O_2$ is also extremely important in the EF process. This cathode type can also be used in the AO process known as AO-$H_2O_2$. Carbonaceous materials such as carbon nanotubes [43], activated carbon fiber [44], and carbon felt (CF) [45] are efficient cathodes for $H_2O_2$ generation. CF is the most widely used cathode due to its high specific surface area and low cost, and it is very efficient in degrading organic pollutants [46-50].

The efficiency of pollutant degradation in the EF process can be increased by exposing the electrochemical cell to light irradiation. In this case, the process is called photoelectro-Fenton (PEF), in which Fe(OH)$^{2+}$ complexes decomposed, leading to the formation of Fe$^{2+}$ and •OH (Eq. 5). Then Fe$^{2+}$ can be reused as a catalyst in the EF process, leading to an increase in •OH formation, thus improving the degradation efficiency of organic pollutants. Furthermore, UV irradiation can also decompose $H_2O_2$ and generate supplementary $^\bullet OH$ (Eq.(6)) [51, 52].

$$Fe(OH)^{2+} + h\nu \rightarrow {^\bullet OH} + Fe^{2+} \tag{5}$$

$$H_2O_2 + h\nu \rightarrow 2\ {^\bullet OH} \tag{6}$$

Bai *et al*. [53] reported an excellent removal efficiency of ciprofloxacin using PEF compared to photocatalysis and EF, achieving 100% elimination within 2 h under visible light irradiation. Titchou *et al*.[48] studied the degradation of direct red 23 by AO, EF, photo-anodic oxidation, and PEF, highlighting the total removal of Direct Red 2 after 6 h of electrolysis by the PEF process.

Gamarra-Güere *et al.* [48] investigated MP degradation using Fenton, photo-Fenton, EF, and electrooxidation processes. The EF process using a $Ti/Ru_{0.3}Ti_{0.7}O_2$ cathode and Pt anode resulted in higher methylparaben degradation. Steter *et al.* [54] investigated MP oxidation by electro-oxidation, EF, and PEF processes using a carbon-PTFE air-diffusion cathode and BDD, Pt, and DSA anodes. Higher mineralization was achieved using the BDD anode, and mineralization efficiency followed PEF > EF > electro-oxidation. Steter *et al.* [55] described the electrochemical degradation of methylparaben using a boron-doped diamond anode alone and/or coupled to sonolysis. The hybrid process achieved 50% removal of total organic carbon (TOC). Fortunato et al. [54] used a BDD anode, and a gas diffusion cathode configuration based on Pd-modified Printex L6 carbon in another study. Pollutant degradation followed pseudo-first-order reaction kinetics in the following order: anodic oxidation coupled with $H_2O_2$ generation (AO- $H_2O_2$) < AO- $H_2O_2$/UVC < electro-Fenton (EF) < photoelectro-Fenton (PEF).

The studies described above use cathodes such as Ti/Ru$_{0.3}$Ti$_{0.7}$O$_2$, carbon-PTFE air-diffusion, and gas diffusion cathode based on Pd-modified Printex L6 carbon. These cathodes may have a higher cost than the carbon felt cathode used in this article. In addition, gas diffusion electrodes require a system for inserting O$_2$ gas, in addition to the addition of PTFE, which can increase the cost of this electrode. The carbon felt, which is used as a cathode in this article, does not have any addition of PTFE or metals, in addition to not being a gas diffusion type cathode, not requiring a system for direct gas insertion. These differences make the felt cathode an electrode that is easy to handle and has a lower cost than those described above.

The present study explored and compared for the first time the use of carbon felt as a cathode in different EAOPs, such as AO-H$_2$O$_2$, EF, and PEF for the removal of MP from aqueous media along with BDD anode. The effect of current density on MP degradation and mineralization was evaluated in all processes used. Subsequently, the AO, EF, and PEF processes were compared in terms of TOC removal efficiency, mineralization current efficiency (MCE%), and energy consumption (EC). Furthermore, the second-order rate constant for the oxidation of MP by hydroxyl radicals was determined using the competition kinetics method. In addition, the intermediate species formed during the MP oxidation were monitored by high-performance liquid chromatography. Based on the data obtained, we proposed a plausible mineralization pathway for the mineralization of MP by hydroxyl radicals.

## 2. Material and methods

*2.1. Chemicals*

MP ($C_8H_8O_3$) with > 98% purity was purchased from Sigma-Aldrich. Sigma-Aldrich also supplied Acetonitrile and sulfuric acid. Sodium sulfate ($Na_2SO_4 \geq 99\%$ purity) and iron (II) sulfate heptahydrate ($FeSO_4\ 7H_2O \geq 99.5\%$ purity) were purchased from Thermo Scientific and Acros Organics, respectively.

The following reagents were also used: 4-nitrophenol (Acros Organics), oxalic acid (Fluka), malic acid (Fluka), succinic acid (Acros Organics), formic acid (Sigma-Aldrich), malonic acid (Fluka), 4-hydroxybenzoic acid (Acros Organics), hydroquinone (Sigma-Aldrich), benzoquinone (Prolabo), and 1-2-4 benzenetriol (Sigma-Aldrich).

*2.2. Electrolytic System*

Homogeneous AO-$H_2O_2$, EF, and PEF experiments were conducted at room temperature in a 200 mL undivided cylindrical glass cell with magnetic stirring and pressurized air bubbling. Bubbling started 5 min before the experiments to ensure saturated oxygen conditions in the solution, essential for generating $H_2O_2$. A BDD film anode on niobium support (CONDIAS GmbH, Itzehoe, Germany) and an unmodified carbon felt cathode (MERSEN, Paris, France) were used, both with a surface area of $2 \times 2\ cm^2$, separated by about 2 cm. Constant-current electrolysis conditions ranged from 10 to 75 mA in a 100 mL aqueous solution containing 0.1 mM MP as a pollutant in 50 mM $Na_2SO_4$ solution [56]. For the EF and PEF processes, the solution pH was adjusted to 3, considered the optimal value for EF and related processes [56], and 0.1 mM $FeSO_4$ was used as the catalyst source. A Rohde Schwarz Hames HM7042-5 triple power supply powered the electrochemical cell. For the PEF process, a 200 W medium-pressure mercury arc lamp (Oriel 6137) emitting a spectral distribution of 254-579 nm was employed as the light source, and the temperature of the electrolyte solution was maintained at 20 degrees. Samples were collected during electrolysis at predefined time intervals to evaluate the concentration of MP and its oxidation products, following the decay kinetics of MP, mineralization rate, and evolution of byproducts.

*2.3. Analytical Procedures*

All MP concentrations were determined using high-performance liquid chromatography (Thermo Scientific Vanquish with a Hypersil GOLD column (250 mm × 4 mm) at 40 °C, with a flow rate of 0.4 mL min$^{-1}$, operated at 254 nm and a mobile phase of acetonitrile/water 40:60. The injection volume was 50 µL. To identify and quantify byproducts observed in the chromatograms obtained during MP degradation, possible byproducts reported in the literature [54, 57, 58] were injected under the same experimental conditions used for MP degradation by comparing retention time and peak area of standard solutions.

The extent of mineralization during MP treatment was monitored by reducing TOC value using a Shimadzu TOC-VCSH analyzer according to the thermal catalytic oxidation principle. The carrier gas was oxygen with a flow rate of 150 mL min$^{-1}$. The oven temperature was 680 °C. Pt was used as a catalyst for the total combustion reaction at this temperature. Calibration of the analyzer was performed with a potassium hydrogen phthalate standard. Reproducible TOC values were obtained using the non-purgeable organic carbon method with an accuracy of ± 2% by injecting 3 mL aliquots into the analyzer.

The evolution of short-chain carboxylic acids generated during MP degradation was carried out using a Merck Lachrom liquid chromatography system equipped with a quaternary pump L-7100 and with a Rezex column (300 mm, 78 mm) coupled to a UV detector set at 220 nm. 0.01 mM $H_2SO_4$ was used as the mobile phase in isocratic elution mode at a flow rate of 0.7 mL min$^{-1}$. The standard solution's retention time and peak area were compared to identify and quantify short-chain carboxylic acids.

## 3. Results and discussion

### *3.1. Effect of current on the removal of MP by AO-H$_2$O$_2$ and EF*

The applied current is one of the most essential parameters in electrochemical processes; it affects the amount of hydroxyl radicals generated in EAOPs through Eqs. (1-4). Fig. 1 shows the effect of current density on the MP concentration decay kinetics using carbon felt as a cathode and BDD as an anode in the EF process (Fig. 1 (a)) and AO-H$_2$O$_2$ process (Fig. 1(b)). It can be noted that the total degradation of 0.1 mM of MP was achieved within 20 min in the EF process with a current of 5 mA cm$^{-2}$. In contrast, the AO-H$_2$O$_2$ process achieved a maximum degradation of 62 % in 60 min, even at a higher current of 10 mA cm$^{-2}$. This rapid degradation of MP in the EF process compared to AO-H$_2$O$_2$ is due to the additional generation of large amounts of homogeneous •OH via the Fenton reaction (Eq. (2)) [51].

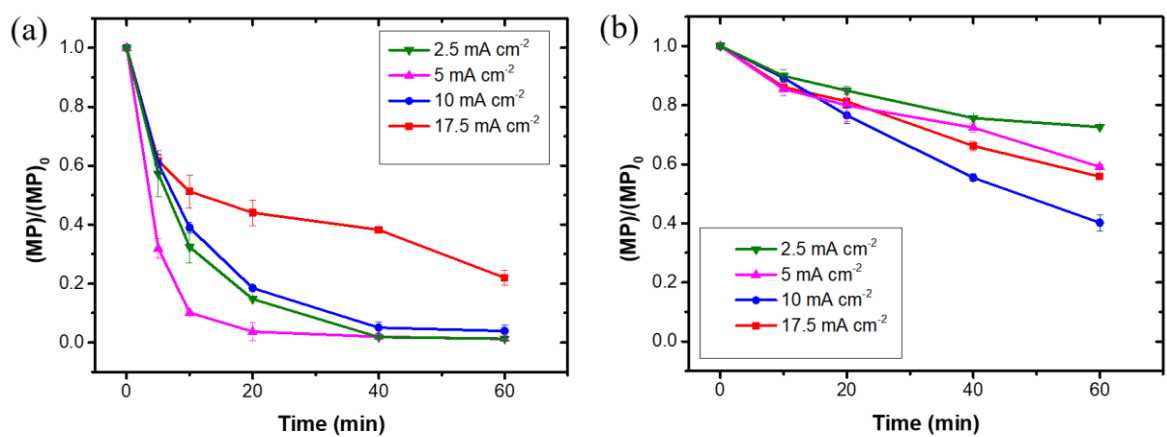

**Figure 1.** Effect of current density on 0.1 mM MP concentration decay during the EF (a) and AO-H$_2$O$_2$ (b) processes in 50 mM Na$_2$SO$_4$ using CF as cathode and BDD as an anode. The catalyst (Fe$^{2+}$) concentration for EF process was 0.1 mM, and the solution pH was 3.

To compare the effect of current density in each process, a kinetic analysis was performed, considering a pseudo-first-order reaction, according to Eq.(7) [49]:

$$\ln\left(\frac{[MP]_0}{[MP]_t}\right) = k_{app} \times t \qquad (7)$$

where $k_{app}$ is the apparent rate constant, and t is the electrolysis time. The slope of the plot ln[MP]$_0$/[MP] versus t (Figs. S1a, S1b) provides the $k_{app}$ values depicted in Table 1.

**Table 1.** Apparent rate constants ($k_{app}$) for MP (0.1 mM) degradation by hydroxyl radicals during EF and AO-H$_2$O$_2$ processes at different current densities in 50 mM Na$_2$SO$_4$ using CF as cathode and BDD as anode. The EF process was performed at pH 3 with a Fe$^{2+}$ concentration of 0.1 mM.

| Process | Current density (mA cm$^{-2}$) | $k_{app}$ (min$^{-1}$) | R$^2$ |
|---|---|---|---|
| AO-H$_2$O$_2$ | 2.5 | 0.007 | 0.998 |
| AO-H$_2$O$_2$ | 5 | 0.008 | 0.986 |
| AO-H$_2$O$_2$ | 10 | **0.015** | 0.973 |
| AO-H$_2$O$_2$ | 17.5 | 0.011 | 0.989 |
| EF | 2.5 | 0.102 | 0.995 |
| EF | 5 | **0.232** | 0.998 |
| EF | 10 | 0.091 | 0.999 |
| EF | 17.5 | 0.059 | 0.810 |

It can be seen from the $R^2$ values that the decay in MP concentration obeys the pseudo-first kinetic model for EF and AO-$H_2O_2$ processes. The EF process provided higher $k_{app}$ values for all current densities than the AO process. Furthermore, the MP concentration decay was improved by increasing the current density to 5 mA cm$^{-2}$ for the EF process and up to 10 mA cm$^{-2}$ for AO- $H_2O_2$. However, a further increase in current density decreased MP degradation kinetics for both methods. This behavior can be explained by the increased rate of side reactions that inhibit the formation of •OH. Higher currents favor the $O_2$ evolution reaction at the anode and the $H_2$ evolution reaction at the cathode, decreasing the production efficiency of M(•OH) at the anode and of electrogeneration of $H_2O_2$ at the cathode, which consequently reduces the number of hydroxyl radicals generated in the system [36, 48, 59].

The absolute rate constant for the second-order kinetics of the reaction between MP and •OH was then determined using the competition kinetics method, which is based on the competitive degradation of the target molecule and a standard competitor with a well-known rate constant [40, 60]. The 4-nitrophenol (NTP), with an absolute rate constant of 3.8 x 10$^9$ M$^{-1}$ s$^{-1}$ [61] was used as a standard competitor under the following experimental conditions: BDD/carbon felt cell, [MP]$_0$ = 0.1 mM, pH = 4, and current of 5 mA cm$^{-2}$. A lower current density value was chosen to avoid interference from the formed intermediates, and pH 4 was selected for the complete solubilization of NTP. The absolute rate constant was then determined according to Eq. (10) [49],

$$\ln\left(\frac{[MP]_0}{[MP]_t}\right) = \frac{k_{MP}}{k_{NTP}} \ln\left(\frac{[NTP]_0}{[NTP]_t}\right) \tag{8}$$

Where $k_{MP}$ and $k_{NTP}$ are the absolute rate constants for MP and NTP hydroxyl radical oxidation reactions, respectively. The slope of the straight line in Fig. 2 allowed the determination of $k_{MP}$ as $8.0 \times 10^8$ M$^{-1}$ s$^{-1}$.

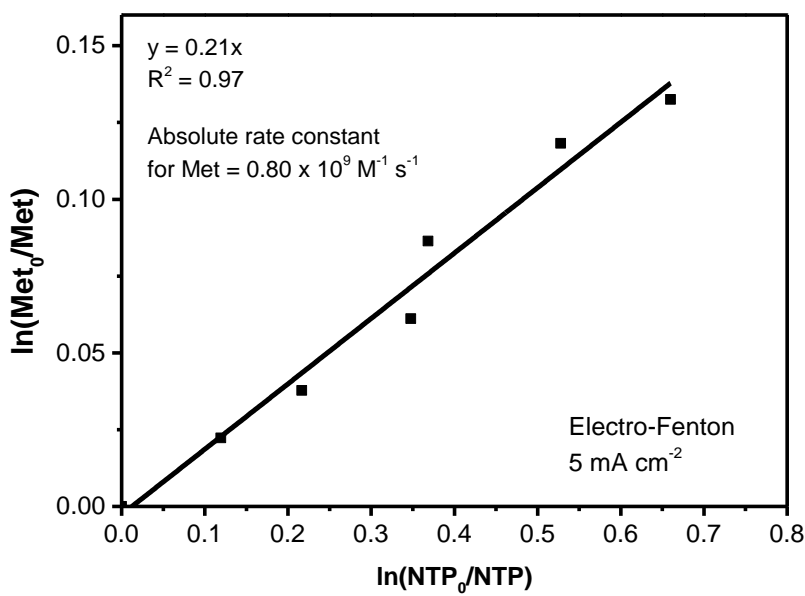

**Figure 2.** Determination of the absolute rate constant for oxidation of MP (0.1 mM) by hydroxyl radicals using competition kinetics method EF process at 5 mA cm$^{-2}$. Experimental conditions: pH = 4, [Na$_2$SO$_4$] = 50 mM, [Fe$^{2+}$] = 0.1 mM, I = 5 mA cm$^{-2}$, CF as cathode, and BDD as an anode.

### *3.2. Effect of current on the mineralization of MP*

Mineralization of MP (0.1 mM) was performed by the EF and AO-H$_2$O$_2$ processes at different applied current values, and results are presented in Fig. 3 in terms of TOC removal percentage as a function of treatment time for the EF (Fig. 3a) and AO-H$_2$O$_2$ (Fig. b) processes. It can be noted that the EF process achieved 85% and 92% TOC removal in 4 and 6 h of treatment, respectively, under a current density of 10 mA cm$^{-2}$. The AO-H$_2$O$_2$ process achieved a maximum TOC removal of 63% and 75% after 4 and 6 h of treatment, respectively, at the same current. The higher TOC removal rates and the greater degradation efficiencies observed previously in the EF process prove their greater efficiency than AO-H$_2$O$_2$. The generation of homogeneous •OH and the heterogeneous M(•OH) produced at the anode makes the EF process more efficient in MP mineralization. Both processes achieved the highest removal rates at higher applied currents because of generating a high amount of M(•OH) at the anode (Eq. 1) and •OH radicals in the bulk solution through Eqs. 2-4 [62].

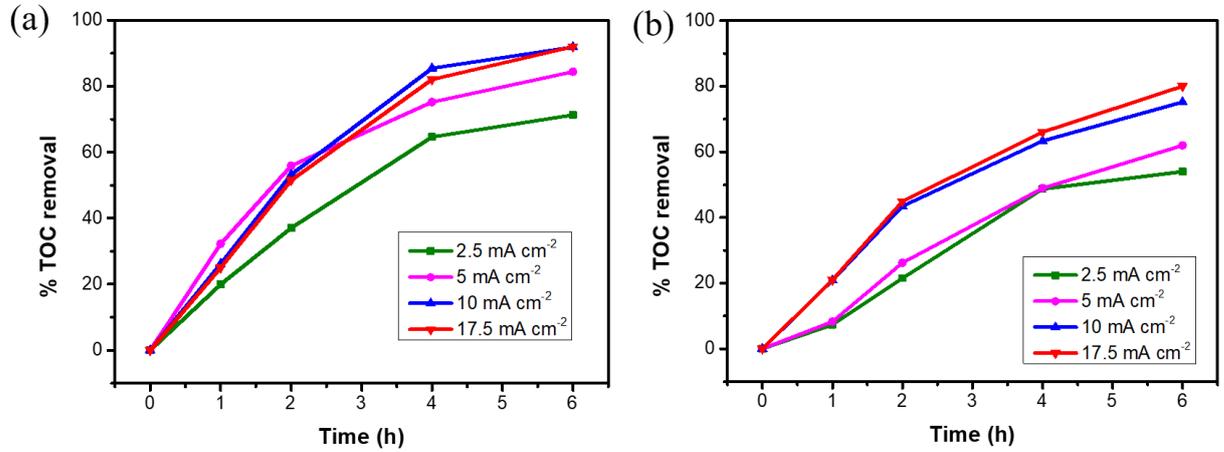

**Figure 3.** Effect of current on TOC removal rate during mineralization of MP solution (0.1 mM) by the EF (a) and AO-$H_2O_2$ (b) processes in 50 mM $Na_2SO_4$ using CF as cathode and BDD as anode. The EF process was performed at pH 3 with a $Fe^{2+}$ concentration of 0.1 mM.

It can be observed that the difference in TOC removal percentage is low for current densities of 10 and 17.5 mA cm$^{-2}$ for both processes. This behavior is generally observed in EAOPs and can be explained by the enhancement of side reactions for high currents, such as the evolution of $O_2$ at the anode (Eq. (9)) and oxygen reduction reaction through the 4-electron at the cathode (Eq. (10)), which prevents the generation of $H_2O_2$. Furthermore, high currents can also favor the rate of waste reactions, such as reduction of $H_2O_2$ at the cathode (Eq. (11)) and dimerization M(•OH) (Eq. (12)) [63, 64].

$$2H_2O \rightarrow O_2 + 4H^+ + 4e^- \qquad (9)$$

$$O_2 + 4H^+ + 4e^- \rightarrow 2H_2O \qquad (10)$$

$$H_2O_2 + 2H^+ + 2e^- \rightarrow 2H_2O \qquad (11)$$

$$2M(•OH) \rightarrow 2M + H_2O_2 \qquad (12)$$

## 3.3. Mineralization current efficiency and energy consumption

The mineralization current efficiency (MCE) was calculated for the AO-$H_2O_2$ and EF processes from Eq. (13) [36],

$$MCE(\%) = \frac{n\, F\, V_S\, \Delta(TOC)_{exp}}{4.32 \times 10^7\, m\, I\, t} \, 100 \qquad (13)$$

where F is the Faraday constant (96485 C mol$^{-1}$), $V_s$ is the solution volume (L), $\Delta(TOC)_{exp}$ is the experimental TOC decay (mg L$^{-1}$), $4.32 \times 10^7$ is the conversion factor (3600 s h$^{-1}$ × 12,000 mg C mol$^{-1}$), m is the number of carbon atoms of MP molecule (m = 8), I is applied current (A) and t is the electrolysis time (h). The number of electrons (n) for the mineralization of mol of MP was taken as 34 considering the theoretical mineralization reaction of the MP (Eq.(14)):[54]

$$C_8H_8O_3 + 13\, H_2O \rightarrow 8\, CO_2 + 34\, H^+ + 34\, e^- \qquad (14)$$

Fig. 4 presents the calculations obtained for the EF (a) and AO-$H_2O_2$ (b) processes. It can be noted that higher MCE values were obtained for the EF process at all applied currents, thereby reinforcing the efficiency of the EF process compared to the AO-$H_2O_2$. The MCE had a slight variation for the AO-$H_2O_2$ compared to the EF due to the low MP mineralization in the former process. In the EF process, MEC progressively decreases over time, which may be related to the decrease of organic matter in the solution and the production of by-products that are difficult to oxidize, such as short-chain carboxylic acids. The MCE values after 6 h of treatment were higher for the current density of 2.5 mA cm$^{-2}$ for both processes and decreased with the current increase. These results may be associated with favoring parasitic reactions at high currents, as mentioned above through Eqs. (9) – (12) [65, 66].

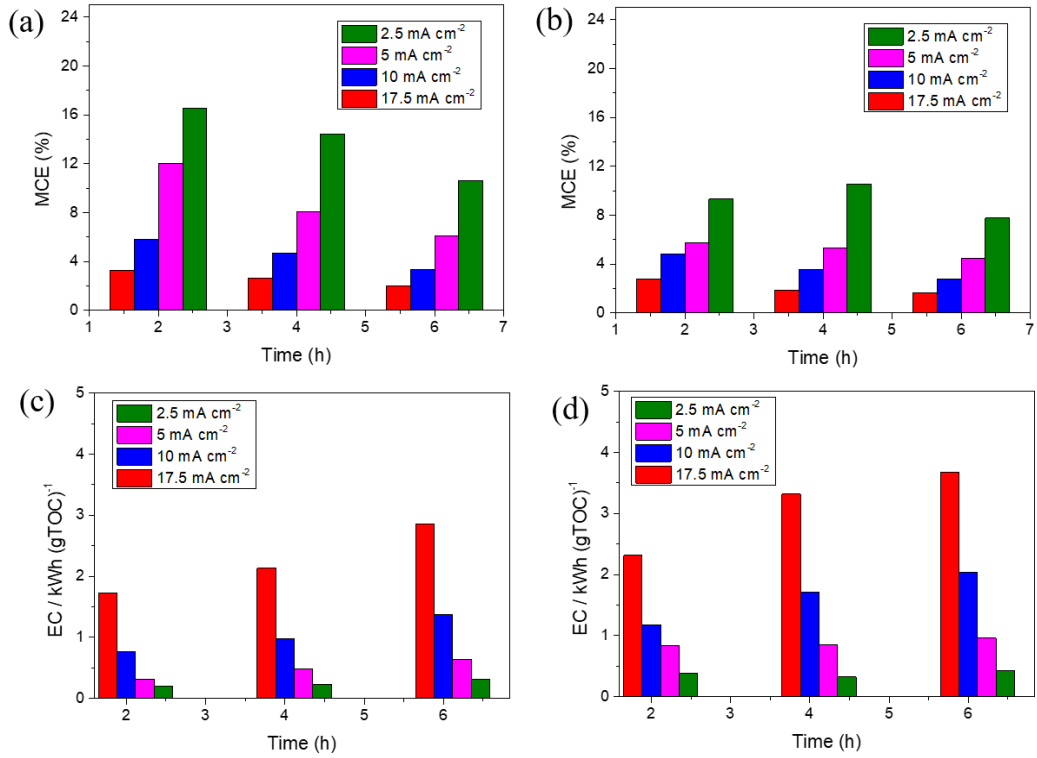

**Figure 4.** Evolution of the MCE over time for the EF (a) and AO-H$_2$O$_2$ (b) processes and evolution of EC with electrolysis time for the EF (c) and AO-H$_2$O$_2$ (d) processes in 50 mM Na$_2$SO$_4$ and 0.1 mM MP using CF as cathode and BDD as anode. The EF process was performed at pH 3 with a Fe$^{2+}$ concentration of 0.1 mM.

The energy consumption (EC) in kWh per g TOC removed for AO-H$_2$O$_2$ and EF processes was also calculated using Eq. (15) [36]:

$$EC(kWh(gTOC)^{-1}) = \frac{E_{cell} \, I \, t}{V_S \, \Delta(TOC)_{exp}} \quad (15)$$

where E$_{cell}$ is the mean potential difference of the cell (V), other terms are identical as defined for Eq. (14). Fig. 4 shows the EC results for the EF (c) and AO-H$_2$O$_2$ (d) processes. It can be noted that the AO-H$_2$O$_2$ presented higher EC values than the EF process at all current density values, which is in agreement with its lower MCE values compared to the EF process [67]. Furthermore, higher current densities and longer electrolysis times lead to a higher CE for both processes.

Table 2 summarizes the TOC removal, EC, and MCE (%) results after 6 h of treatment. The EF process stands out at all current densities and is considered more efficient than the AO-$H_2O_2$ process for MP degradation. Analyzing the results of the EF process in more depth, the current density of 10 mA cm$^{-2}$ better results, achieving a TOC removal of 92% at a medium cost.

**Table 2.** % TOC removal, MCE (%), and EC values for the EF and AO-$H_2O_2$ processes at different current densities after 6 hours of treatment.

| Process | Current density (mA cm$^{-2}$) | % TOC removal | MCE(%) | EC (kWh(gTOC)$^{-1}$) |
|---|---|---|---|---|
| AO-$H_2O_2$ | 2.5 | 53.85 | 7.75 | 0.43 |
| AO-$H_2O_2$ | 5 | 61.51 | 4.50 | 0.96 |
| AO-$H_2O_2$ | 10 | 74.98 | 2.79 | 2.04 |
| AO-$H_2O_2$ | 17.5 | 79.86 | 1.63 | 3.67 |
| EF | 2.5 | 80.00 | 10.61 | 0.32 |
| EF | 5 | 85.15 | 6.06 | 0.64 |
| EF | 10 | 91.86 | 3.35 | 1.38 |
| EF | 17.5 | 92 | 1.99 | 2.86 |

### 3.4. Comparison of the EF and photoelectro-Fenton processes

As reported above, the EF process showed more significant MP degradation and mineralization efficiency than the AO-$H_2O_2$ process. Therefore, the EF process was further explored to improve its efficiency by combining it with UV light irradiation, called photoelectron (PEF). This coupling constitutes an efficient strategy to increase the removal of organic pollutants [68] and was explored for the degradation of MP under the same experimental conditions as the EF process, except for UV light.

Fig. 5 shows the effect of current density on the MP concentration decay kinetics using EF and PEF processes. It can be noted that the degradation by the PEF process was improved compared to the EF process at all current densities studied. During the degradation of organic pollutants by the EF process, the generated carboxylic acids form stable complexes with $Fe^{3+}$ and decrease mineralization efficiency. The UV irradiation destroys these complexes, regenerating the catalyst ($Fe^{2+}$) and supplementary $^{\bullet}OH$ (Eq. (16)), in addition to the photolysis of $H_2O_2$ to generate $^{\bullet}OH$ (Eq. (6)) [48, 51].

$$Fe(OOCR)^{2+} + h\nu \rightarrow Fe^{2+} + CO_2 + {}^{\bullet}OH \tag{16}$$

The $k_{app}$ values, obtained by the slope of the graph $\ln[MP]_0/[MP]_t$ versus t (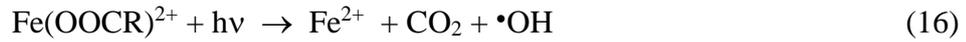Fig. S2), were calculated to compare EF and PEF processes in terms of the degradation rate of MP (Table 3). The $k_{app}$ values for PEF are higher than those observed for the EF process, given in Table 1. However, this difference is more noticeable at higher current densities. The $k_{app}$ values increased from 0.091 to 0.155 min$^{-1}$ at 10 mA cm$^{-2}$ and from 0.059 to 0.069 min$^{-1}$ at 17.5 mA cm$^{-2}$. This behavior may be associated with the higher production of $^{\bullet}OH/M(^{\bullet}OH)$ and the regeneration of $Fe^{2+}$. Furthermore, it can be observed that a further increase in current density leads to a decrease in $k_{app}$ values. This result is due to a rise in the rate of side reactions that inhibit the formation of $^{\bullet}OH$, decreasing the $k_{app}$ values, as previously explained.

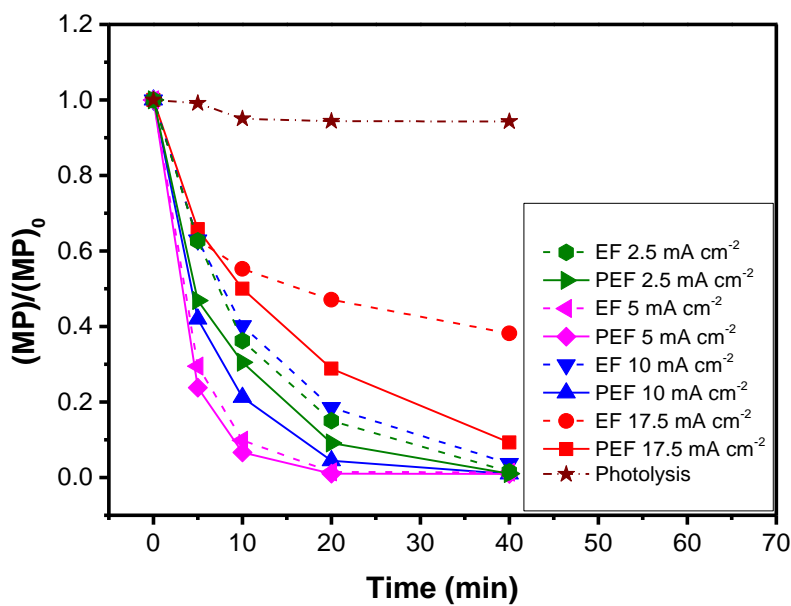

**Figure 5.** Effect of current density on MP concentration decay (0.1 mM) by EF and PEF processes in 50 mM $Na_2SO_4$, $[Fe^{2+}]$ = 0.1 mM, pH = 3, and using CF as cathode and BDD as anode (a).

**Table 3.** Apparent rate constants ($k_{app}$) for MP (0.1 mM) degradation by PEF at different current densities in 50 mM $Na_2SO_4$, pH 3 with a $Fe^{2+}$ concentration of 0.1 mM, and using CF as the cathode and BDD as the anode.

| Current density (mA cm$^{-2}$) | $k_{app}$ (min$^{-1}$) | $R^2$ |
|---|---|---|
| 2.5 | 0.119 | 0.950 |
| 5 | 0.271 | 0.997 |
| 10 | 0.155 | 0.998 |
| 17.5 | 0.069 | 0.971 |

For a more in-depth comparison between EF and PEF, the effect of current density was also investigated on the TOC removal and MCE, and results are presented in Fig. 6 (a) and 6 (b), respectively. It can be noted that the PEF process provides better TOC removal rates compared to the EF process at all applied currents. This reinforces the contribution of •OH formed by UV light irradiation of the solution. The percentage of TOC removal reached 79.9, 84.7, 85.7, and 77.3 % for current densities of 2.5, 5, 10, and 17.5 mA cm$^{-2}$, respectively, in 2 h of treatment. As can be seen in Table 4, these results are significantly higher than those obtained for the EF process for the same treatment time. As previously discussed, the TOC removal degree decreased for the current density of 17.5 mA cm-2, which may be related to the effect of side reactions at high currents.

Analyzing the MCE results presented in Fig. 6(b), the highest values were obtained at lower currents; the same behavior was observed for the EF and AO-H$_2$O$_2$ processes. As explained previously, higher current densities increase parasitic reaction rates, decreasing the MCE.

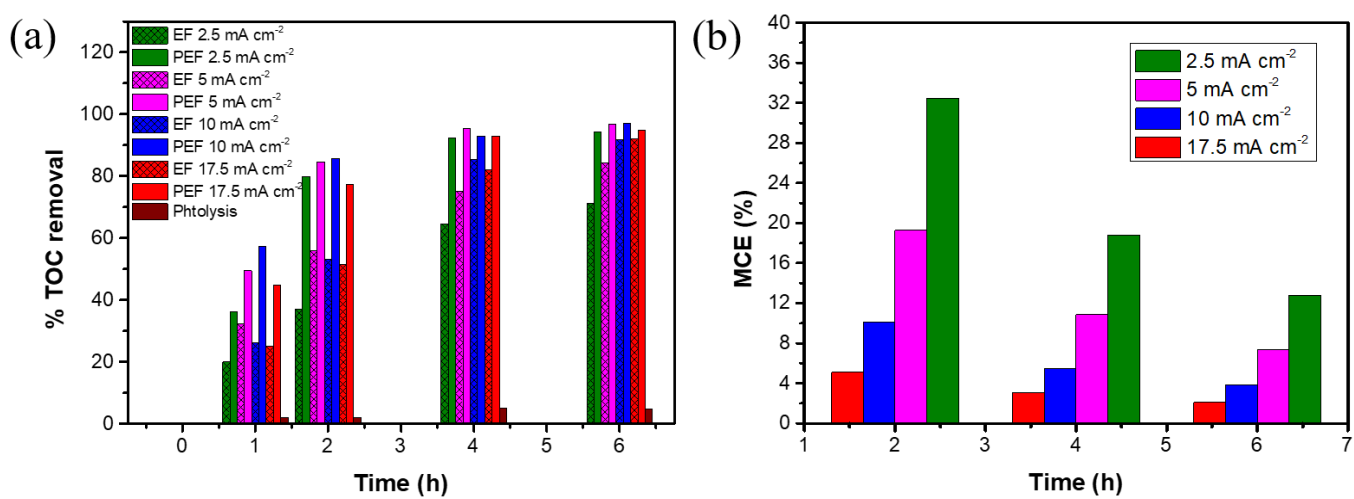

**Figure 6**. Effect of current density on TOC removal rate of MP (0.1 mM) for EF and PEF processes (a) and evolution of the MCE (%) over time for the PEF (b). Experimental conditions: 50 mM Na$_2$SO$_4$, [Fe$^{2+}$] = 0.1 mM, pH = 3, and CF as cathode and BDD as anode

Table 4 shows the MCE results for the EF and PEF processes after 2 h of treatment when the difference between the processes is most relevant. The PEF process presented higher MCE values, the maximum value being 32.45% for the current of 2.5 mA cm$^{-2}$.

**Table 4.** % TOC removal and MCE (%) for the EF and PEF processes at different current densities after 2 hours of treatment.

| Process | Current density (mA cm$^{-2}$) | % TOC removal | MCE(%) |
|---|---|---|---|
| EF | 2.5 | 37.21 | 16.53 |
| EF | 5 | 56.00 | 11.95 |
| EF | 10 | 52.75 | 5.84 |
| EF | 17.5 | 51.29 | 3.33 |
| PEF | 2.5 | 79.89 | 32.45 |
| PEF | 5 | 84.65 | 19.23 |
| PEF | 10 | 85.69 | 10.00 |
| PEF | 17.5 | 77.27 | 5.10 |

The results obtained in this study highlight the efficiency of the PEF process for removing MP from water using a carbon felt cathode and a BDD anode. Other studies on MP degradation are depicted in Table 5. Using UVA light, Steter *et al*. [54] used a carbon-PTFE air-diffusion cathode and BDD anode for MP degradation via the PEF process. Degradation of 158 mg L$^{-1}$ MP solution using a current density of 66.7 mA cm$^{-2}$ achieved 97% TOC removal in 6 h of treatment. Benitez *et al*.[69] achieved 60 % TOC removal from MP (1 mg L$^{-1}$) in 5 h using the photo-Fenton (PF) process with a Xenon lamp. In this way, the degradation of MP by the PEF process using a low-cost material seems to be an efficient strategy for the TOC removal of 84.65 % in 2 h at a low current of 5 mA cm$^{-2}$.

Compared to the other processes depicted in Table 5, carbon felt as a cathode in the EF processes is very efficient. In addition, the carbon-felt cathode is low-cost and does not require other experimental steps, such as gas diffusion electrodes.

**Table 5.** A comparison of different advanced oxidation processes for methylparaben degradation was reported in the literature.

| Process | Experimental conditions | $k_{app}$ | TOC removal | Ref |
|---|---|---|---|---|
| PEF | Carbon-PTFE air-diffusion cathode and BDD anode; 0.025 M $Na_2SO_4$ + 0.035 M NaCl; [MP] 158 mg/L; $j$: 66.7 mA $cm^{-2}$; pH: 3; [$Fe^{2+}$]: 0.5 mmol; Light source: UVA 5 W $m^{-2}$ | - | 97% (360 min) | [54] |
| EO | Ti cathode and BDD anode; 0.05 M $K_2SO_4$; [MP] 100 mg/L; $j$: 10.8 mA $cm^{-2}$ | 0.0029 $min^{-1}$ | 37.76% (120 min) | [70] |
| PF | [MP] 1mg/L; [$Fe^{2+}$]: 21 mg/L; $H_2O_2$: 155.37 mg/L; pH: 3.0; Light source: xenon lamp 350 W/$m^2$ | 0.0686 $min^{-1}$ | 60% (300 min) | [69] |
| Fenton | [MP] 100 mg/L; [$Fe^{2+}$]: 16 mg/L; [$H_2O_2$]: 62 mg/L; pH: 3 | 0.26 $min^{-1}$ | 33.3% (20 min) | [58] |
| PF | [MP] 100 mg/L; [$Fe^{2+}$]: 4 mg/L; [$H_2O_2$]: 52 mg/L; pH 3; Light source: UVC Lamp 4 W | 0.34 $min^{-1}$ | 34.9% (20 min) | [58] |
| EF | DSA Cathode and Pt anode; [MP] 100 mg/L; [$Na_2SO_4$]: 0.05 mol/L; [$Fe^{2+}$]: 4 mg/L; $j$: 25 mA $cm^{-2}$; pH: 3 | 0.57 $min^{-1}$ | 28.8% (20 min) | [58] |
| AO-H$_2$O$_2$ | Carbon-felt cathode and BDD anode; 50 mM Na$_2$SO$_4$; [MP] 0.1 mM; $j$: 17.5 mA cm$^{-2}$ | 0.011 min$^{-1}$ | 79.86% (360 min) | This work |
| EF | Carbon-felt cathode and BDD anode; 50 mM Na$_2$SO$_4$; [MP] 0.1 mM; [Fe$^{2+}$]: 0.1 mM; $j$: 17.5 mA cm$^{-2}$; pH:3 | 0.059 min$^{-1}$ | 92% (360 min) | This work |
| PEF | Carbon-felt cathode and BDD anode; 50 mM Na$_2$SO$_4$; [MP] 0.1 mM; [Fe$^{2+}$]: 0.1 mM; $j$: 10 mA cm$^{-2}$; pH:3; Mercure lamp 200 W | 0.155 min$^{-1}$ | 85.69% (120 min) | This work |

### *3.6. Identification and evolution of carboxylic acids and different by-products*

The degradation of organic pollutants by hydroxyl radicals leads to forming a variety of short-chain carboxylic acids and degradation by-products before complete mineralization, i.e., the transformation of organics under study to CO$_2$ and H$_2$O [63]. The byproducts were identified and followed for the EF process at a current density of 10 mA cm$^{-2}$ during 6 h of treatment. The results are presented in Fig. 7. Ion exclusion chromatograms allowed the detection of five carboxylic acids with well-defined peaks corresponding to oxalic, malic, succinic, formic, and malonic acids at the retention times ($t_R$) of 6.4, 8.81, 12.10, 13.00, and 10.13 min, respectively. The concentration profile of these carboxylic acids, presented in Fig. 7 (a) shows a rapid accumulation in the initial phase of treatment of succinic and malic acids, which are entirely mineralized after 150 min. Malonic acid appears after 100 min of treatment and remains below 0.005 mM. Formic acid had a maximum concentration of 0.015 mM and was removed entirely in 250 min of treatment. Oxalic acid, considered the final product before complete mineralization [71], was identified in the highest concentration of 0.025 mM and reached a final concentration of 0.007 mM at the end of treatment. These results corroborate with the TOC removal data under the same experimental conditions. The aromatic intermediate products, such as 4-hydroxybenzoic acid, hydroquinone, benzoquinone, and 1,2,4-benzenetriol, presented in Fig. 7 (b), were also identified in the HPLC chromatograms with well-defined peaks at $t_R$ of 6.4, 8.08, 8.1, and 5.7, respectively. All other byproducts were degraded within 100 min of treatment except benzoquinone. 1-2-4 benzenetriol was present at a concentration of $2.37 \times 10^{-7}$ mM in the solution even after 6 h of treatment. Oxalic acid and benzoquinone form the residual TOC at the end of the treatment.

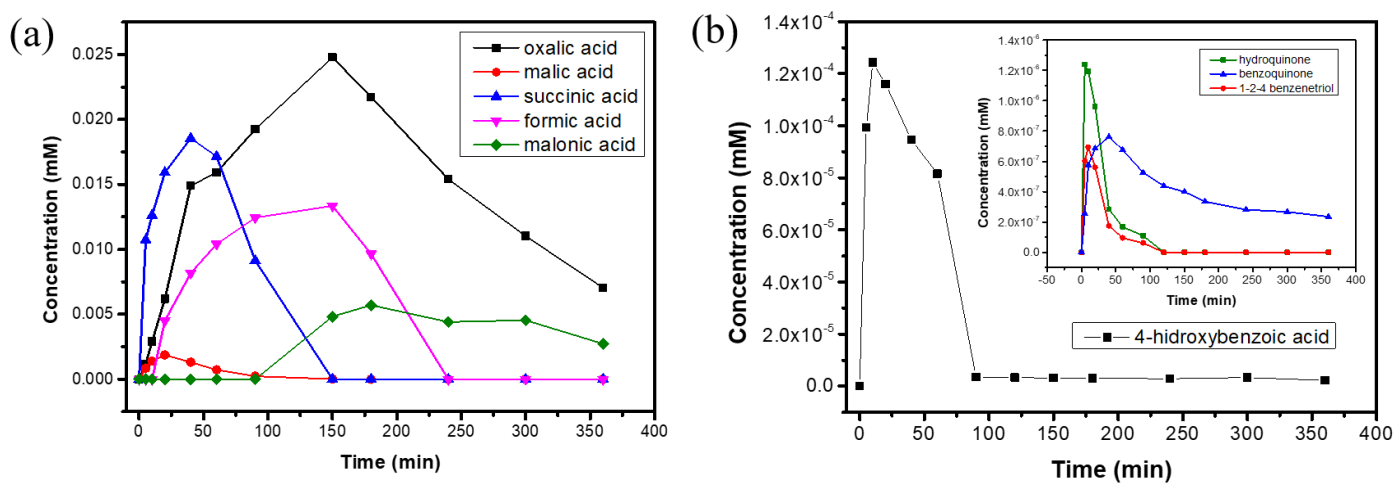

**Figure 7.** Evolution of short-chain carboxylic acids (a) and other byproducts during the EF treatment of 0.1 mM MP with a current density of 10 mA cm$^{-2}$, in 50 mM Na$_2$SO$_4$ medium. [Fe$^{2+}$] = 0.1 mM, pH = 3, CF as cathode and BDD as anode.

### *3.7. Mineralization Pathway for MP*

A mineralization route for MP mineralization was proposed considering only the contribution of homogeneous (•OH) and heterogeneous BDD(•OH) hydroxyl radicals. Degradation was initiated by the hydroxylation of MP (I), leading to the formation of 4-hydroxybenzoic acid (II), which is then hydroxylated in its turn to hydroquinone (III).

Then, 1,2,4-benzenetriol (IV) [72] is formed by hydroxylation of hydroquinone. Benzoquinone (V) can be formed by hydroxylation or electrochemical oxidation at the anode. The oxidation of these intermediates (III, IV, V) causes the opening of the ring of these aromatic structures, producing short-chain carboxylic acids such as succinic (VI), malic (VII), malonic (VIII), oxalic (IX), and formic (X) acids. The oxidation of carboxylic acids to $CO_2$ and water constitutes the last stage of mineralization according to the TOC removal values [57, 70].

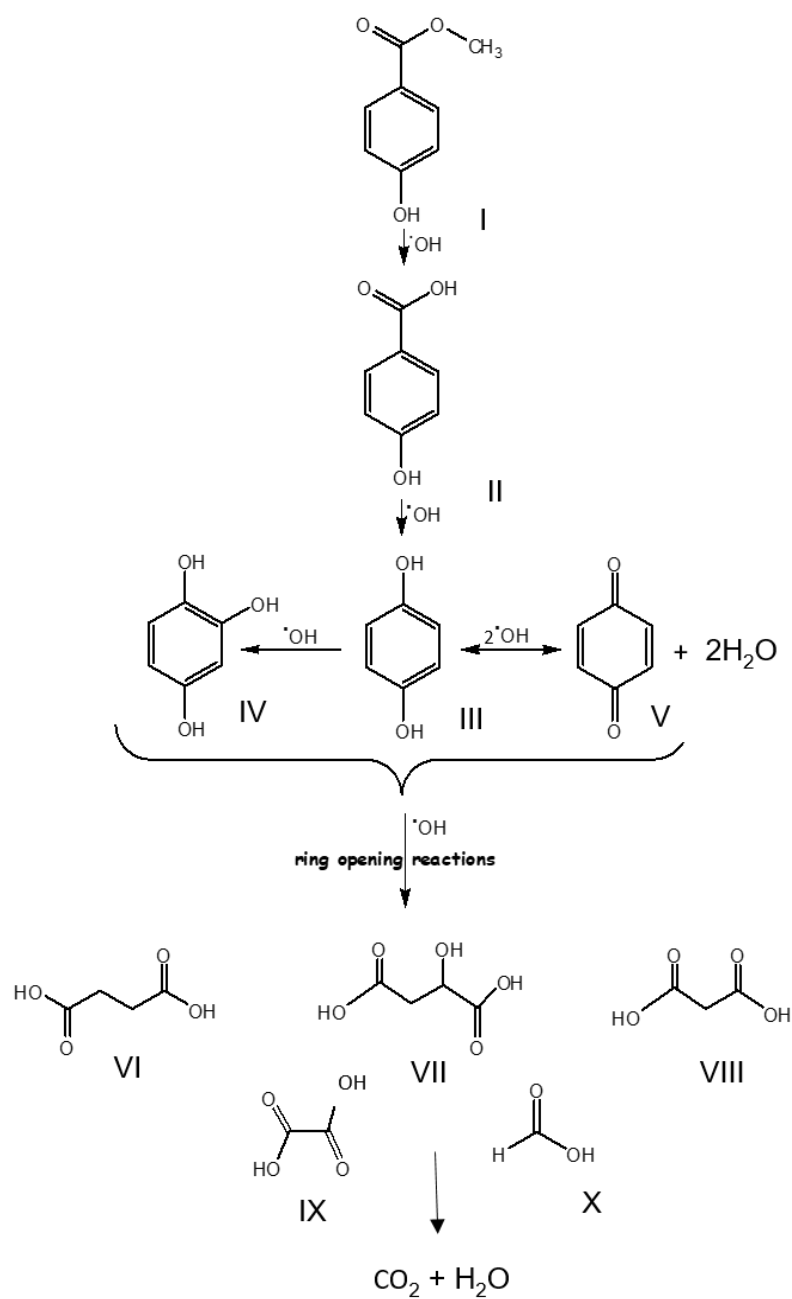

**Figure 8.** Reaction sequence for MP mineralization by hydroxyl radicals during the EF process using a carbon felt cathode and a BDD anode.

## 4. Conclusions

The oxidative degradation of methylparaben (MP) and the mineralization of its 0.1 mM solution were studied using various electrochemical advanced oxidation processes (EAOPs), including AO-$H_2O_2$, EF, and PEF, under different current densities with a carbon felt cathode and a BDD anode. Among these, the EF process demonstrated superior MP removal efficiency compared to the AO-$H_2O_2$ process at all current densities. Higher current densities enhanced the mineralization rate, as reflected by the percentage of TOC removal by both processes. After 6 h of treatment, the TOC removal rates reached 91.86% and 74.98% for the EF and AO-$H_2O_2$ processes, respectively, at a 10 mA cm$^{-2}$ current density. This indicates that the EF process is more efficient and offers higher current efficiency and lower energy consumption, making it the most cost-effective option compared to AO-$H_2O_2$. In ddition, combining EF with UV light in the PEF process improved the removal efficiency of the target pollutants. The PEF process achieved 84.65% TOC removal within 2 hours at a lower current density of 5 mA cm$^{-2}$, significantly outperforming the EF process efficiency, which gained only 55.92% under the same conditions. The kinetics of MP degradation followed a pseudo-first-order reaction trend in the sequence: AO-$H_2O_2$ < EF < PEF.

**Acknowledgments**

The authors are grateful to the following Brazilian research financing institutions: São Paulo Research Foundation (FAPESP grants, #2021/14394-7, #2022/15252-4,

#2022/12895-1, #2023/01187-9, #2023/02396-0, #2021/05364-7) and Conselho Nacional de Desenvolvimento Científico e Tecnológico (CNPq) (#303943/2021-1, #308663/2023–3, #402609/2023–9) for the financial assistance provided in this work.